\documentclass[floatfix,twocolumn,showpacs,preprintnumbers,amsmath,amssymb]{revtex4}

\newcommand{\bib}{\bibitem}
\newcommand{\bea}{\begin{eqnarray}}
\newcommand{\eea}{\end{eqnarray}}
\newcommand{\beq}{\begin{equation}}
\newcommand{\eeq}{\end{equation}}
\newcommand{\non}{\nonumber}

\newcommand{\de}{\delta}

\newcommand{\la}{\lambda}

\newcommand{\si}{\sigma}
\newcommand{\pa}{\partial}

\newcommand{\Mm}{\mathbb{M}}
\newcommand{\Sm}{\mathbb{S}}

\usepackage{graphicx,epsfig} 
\usepackage{dcolumn} 
\usepackage{bm} 

\begin{document}

\title{Power dissipation for systems with junctions of multiple quantum wires}

\author{Amit Agarwal$^1$, Sourin Das$^{2,3}$ and Diptiman Sen$^4$}

\affiliation{$^1$NEST and Scuola Normale Superiore, Piazza dei Cavalieri, 7, 
I-56126 Pisa, Italy \\
$^2$Institut f\"ur Festk\"orper-Forschung -- Theorie 3, Forschungszentrum
J\"ulich, 52425 J{\"u}lich, Germany \\
$^3$Institut f\"ur Theoretische Physik A, RWTH Aachen, 52056 Aachen, Germany \\
$^4$Centre for High Energy Physics, Indian Institute of Science, Bangalore 
560 012, India}

\date{\today}

\begin{abstract}
We study power dissipation for systems of multiple quantum wires meeting at
a junction, in terms of a current splitting matrix ($\Mm$) describing the
junction. We present a unified framework for studying dissipation for wires
with either interacting electrons (i.e., Tomonaga-Luttinger liquid wires with
Fermi liquid leads) or non-interacting electrons. We show that for a given
matrix $\Mm$, the eigenvalues of $\Mm^T \Mm$ characterize the dissipation, 
and the eigenvectors identify the combinations of bias voltages which need 
to be applied to the different wires in order to maximize the dissipation 
associated with the junction. We use our analysis to propose and study some
microscopic models of a dissipative junction which employ the edge states
of a quantum Hall liquid. These models realize some specific forms of the
$\Mm$-matrix whose entries depends on the tunneling amplitudes between the
different edges.

\end{abstract}

\maketitle

\section{Introduction}

One-dimensional systems of strongly correlated electrons have been studied
extensively for several years both experimentally, in the form of quantum
wires and carbon nanotubes, and theoretically 
\cite{gogolin,vondelft,rao,giamarchi1,bruus,giuliani}. 
Junctions of several quantum wires have also been studied in recent years 
since they can now be experimentally realized in carbon nanotubes 
\cite{li,kumar,papad,fuhrer,terrones}. The existing studies of junctions of
quantum wires, which are usually modeled as Tomonaga-Luttinger liquids (TLL),
have mainly looked at their low-temperature fixed points and the
corresponding conductance matrices 
\cite{sandler,nayak,lal1,chen,chamon,meden,das2006,lal2,rosenow,giuliano,bellazzini,das2007,das2008}.
Many of these studies have focused on situations in which there is no power
dissipation in the system. The aim of our work will be to include dissipation
in the discussion. For simplicity, we will consider only spinless electrons
and will restrict ourselves to the zero frequency limit (DC) in our work.
 
A motivation for studying dissipation is as follows. In Ref.
\onlinecite{chamon}, a two-parameter description of a junction of three TLLs
has been discussed. The dissipationless fixed points were shown to lie on the 
circumference of a circle, while the interior of the circle corresponds to 
dissipative junction. In this context, the center of the circle which 
corresponds to the current splitting matrix with all its elements equal to 
$1/3$ is of particular interest as it corresponds to the maximum possible 
dissipation allowed by constraint of current conservation. It therefore seems 
useful to understand dissipative junctions in a general way and to study 
whether any of the points inside the circle correspond 
to fixed points of some renormalization group (RG) equations.
 
The plan of this paper is as follows. In Sec. II, we introduce the idea
of a current splitting matrix $\Mm$ at a junction of $N$ wires and discuss
the cases of both non-interacting and interacting electrons. In Sec. III, we
obtain an expression for the power dissipated in terms of this matrix. The 
measure of the degree of dissipation is then defined in terms of the
eigenvalues of $\Mm^T \Mm$. For two-wire and three-wire junctions, we write
down the most general form of the $\Mm$-matrix allowed by current conservation,
thus providing a complete parametrization of the dissipation at the junction.
In general, an $\Mm$-matrix which respects current conservation can have both
positive and negative matrix elements. As we will show, an $\Sm$-matrix 
describing non-interacting electrons scattering at the junction can be 
related to a matrix $\Mm$ all of whose elements are positive; this relation 
will follow from the assumption that there are no phase correlations between 
electrons coming from different reservoirs which lie far away from the 
junction. But when $\Mm$ has negative elements, such a relation does not 
exist and the matrix then necessarily corresponds to a system of interacting 
electrons.

In Sec. IV, we introduce a simple model involving three patches of the edge
states of a quantum Hall liquid with filling fraction $\nu$. The patches
are taken to be mutually coupled to each other by local electron tunnelings
between three distinct points lying on the three patches with amplitudes
$\si_{ij}$, where $i,j$ denote the patch index. Then a parametrization
of the $\Mm$-matrix
is obtained in terms of the conductance amplitudes $\si_{ij}$. In this way
we obtain interesting dissipationless matrices in the limits $\si_{ij} \to 0$ 
and $\infty$ respectively. These matrices were shown to represent
dual fixed points in the theory of a junction of TLL wires in Ref.
\onlinecite{chamon} using more involved calculations. In Sec. V, we introduce
a more complex model of a junction of three quantum wires in which the
junction consists of a ring-shaped region with edges of its own. Once again,
the matrix $\Mm$ of the entire system can be found in terms of the coupling
of each wire to the ring and the tunneling amplitudes across the
two edges of the ring. Even though such a geometry is complicated,
from an experimental point of view it allows for
easier tunability as far as realizing various types of $\Mm$ matrices is
concerned. In Sec. VI, we make some concluding remarks.

\section{The current splitting matrix}

A junction is a meeting point of $N$ wires each of which has an incoming and
an outgoing mode. Physically, if the junction is made of a material like a 
carbon nanotube, then the incoming and outgoing modes (which carry currents) 
belonging to a single wire are not separated in space. But if these are 
quantum wires made out of the edge states of a quantum Hall liquid 
\cite{linejunction}, then the incoming and outgoing modes are spatially 
separated.
In the following discussion, we will consider a junction of several quantum 
wires, each with two spatially separated chiral current carrying edges, one 
incoming and one outgoing. Each chiral mode (incoming or outgoing) is labeled 
by an index $i$ which runs from $1$ to
$N$ and is parametrized by a coordinate $x$. We will take $x$ to run from
$0$ to $\infty$; the point $x=0$ will be common to all the wires and will
denote the junction. The outgoing currents in the system are related to the
incoming currents by a current splitting matrix $\Mm$ given by
\beq J_{Oi} ~=~ \sum_j ~\Mm_{ij} ~J_{Ij}. \label{eq1} \eeq
Current conservation at the junction therefore implies that each column of
$\Mm$ must add up to $1$. Let us also assume that the incoming
current on wire $i$ is proportional to the applied bias voltage $V_{Ii}$,
with the constant of proportionality being the same for all wires. Then if
all the wires have the same bias voltage, the net outgoing current $J_{Oi}
- J_{Ii}$ on each wire $i$ must vanish which implies that each row
of $\Mm$ must also add up to 1.

For non-interacting electrons, the junction can be described in terms of a
scattering matrix $\Sm$ which provides a linear relation between the incoming
and outgoing electron fields at the junction. Namely, the incoming and outgoing
electron fields, $\psi_{Ii} (x,t)$ and $\psi_{Oi} (x,t)$, are related at all
times $t$ as 
\beq \psi_{Oi} (0,t) ~=~ \sum_j ~\Sm_{ij} ~\psi_{Ij} (0,t). \label{smat} \eeq
Current conservation implies that $\Sm$ must be an $N \times N$ unitary
matrix. Any deviation from the linear boundary condition for the electron 
fields at the junction will imply the existence of local inter-electron 
interactions at the junction even if the electrons in the bulk of the wires 
are left non-interacting. The scattering matrix description can also be used 
to describe electrons which are weakly interacting in the bulk of the wire, by
treating the effects due to interactions perturbatively \cite{lal1,chen,meden}.

Given a scattering matrix $\Sm$ for non-interacting electrons, we will now
see how the elements of the current splitting matrix $\Mm$ can be found. The 
incoming and outgoing currents $J_{Ii}$ and $J_{Oi}$ in wire $i$ are 
proportional to $|\psi_{Ii}|^2$ and $|\psi_{Oi}|^2$ respectively. Eq. 
(\ref{smat}) implies that $|\psi_{Oi}|^2 = \sum_{jk} \Sm_{ij}^* \Sm_{ik} 
\psi_{Ij}^* \psi_{Ik}$. We now assume that there are no phase correlations 
between the incoming electrons on different wires $j$ and $k$ since they are
coming from different reservoirs whose distances from the junction are 
taken to be much larger than the phase coherence length; the absence of such 
phase correlations is crucial for the validity of the Landauer-B\"uttiker 
theory of electronic transport in mesoscopic systems \cite{buttiker1}. Hence 
terms like $\psi_{Ij}^* \psi_{Ik}$ can be set equal to zero if $j \ne k$. We 
thus obtain $|\psi_{Oi}|^2 = \sum_j |\Sm_{ij}|^2 |\psi_{Ij}|^2$. This is of 
the same form as in Eq. (\ref{eq1}) if we identify $M_{ij} = |S_{ij}|^2$.

On the other hand, if we have strongly interacting electrons in one dimension,
then it is natural to use bosonization. The electrons in the wire are then
expressed in terms of free bosonic excitations described by TLL theory, and
the fixed point theory of the junction can be described in terms of a current
splitting matrix $\Mm$ which is obtained by imposing a linear boundary
condition on the incoming and outgoing bosonic fields at the junction
\cite{sandler,nayak,chamon,das2006,lal2,rosenow,giuliano,bellazzini,das2007,das2008}.
One can use free bosonic fields to describe either non-interacting or
interacting electrons in the bulk of the one-dimensional wires depending on
whether the Luttinger parameter $g$ is equal to or not equal to 1. Note that
even when we have $g=1$ (non-interacting electrons) in the bulk of the wire, 
within the bosonization approach 
the current splitting matrix $\Mm$ representing a linear relation between the
incoming and outgoing boson fields at the junction corresponds to the presence
of non-zero inter-electron interaction at the junction. This is because
the boson fields are related to the corresponding electron fields by
a non-linear bosonization identity, $\psi_{I/O} (x) =(1/\sqrt{2 \pi \alpha})~
F_{I/O} e^{i \phi_{I/O}(x)}$
\cite{gogolin,vondelft,rao,giamarchi1,bruus,giuliani}, where $\psi_{I/O}(x)$
are the incoming and outgoing electron fields, $\phi_{I/O}$ are the incoming
and outgoing chiral bosonic fields, and $F_{I/O}$ are the corresponding Klein
factors. Thus there is a subtle difference between using a scattering matrix
$\Sm$ for non-interacting electrons and a current splitting matrix $\Mm$ in
bosonized TLL theory for electrons which are non-interacting ($g=1$) in the
bulk of the wire. In the latter case the $\Mm$-matrix description of the
junction corresponds to an interacting theory of electrons where the
interaction is localized at the junction. Now, if the incoming and outgoing
boson fields are linearly related to each other at the junction, i.e., if
$\phi_{Oi} (x=0,t) = \sum_j \Mm_{ij} \phi_{Ij} (x=0,t)$, then $\Mm$ must be a
real and orthogonal field splitting matrix in order that both the incoming
and outgoing bosonic fields satisfy the canonical commutation relations
\cite{das2006,agarwal1,agarwal2}. Such a description of the junction given
by an orthogonal $\Mm$ represents fixed points of the junction as was shown 
in Ref. \onlinecite{das2006}. The current at any point of wire $i$ is given by
$-(1/2\pi) \pa \phi /\pa t$. Hence we note that the above condition at the 
junction implies that the outgoing and incoming currents also satisfy Eq. 
(\ref{eq1}), i.e., the field splitting $\Mm$ matrix can be taken to be the 
same as the current splitting matrix. Hence in the bosonic formalism, $\Mm$ 
must be an $N \times N$ real and orthogonal matrix each of whose rows and 
columns add up to 1 \cite{bellazzini}.

As we will see later, the orthogonality condition on the $\Mm$ matrix also
renders it dissipationless irrespective of its origin, i.e., this is true
for both a junction of non-interacting electrons described by an $\Sm$ matrix
or a junction of TLL wires described by a field splitting
matrix $\Mm$. Hence, to include dissipation in the analysis we have to relax
the condition of orthogonality on $\Mm$. This also implies that the formalism
of bosonization cannot be used directly since a non-orthogonal $\Mm$ does not
allow the bosonic commutation relations to be satisfied. However, $\Mm$ must
continue to be real since it relates incoming and outgoing currents which are
all real, and each of its rows and columns must add up to 1 as argued before.
Hence the main emphasis of this section is on the fact that various situations
comprising of either non-interacting electrons or interacting electrons, where
the interaction is either localized at the junction or extended all over the
wire, can be described just in terms of a current splitting matrix (Eq.
(\ref{eq1})). We will see in the following section that this information
is enough to characterize the dissipation associated with the junction.

\section{Dissipation}

We will now consider the specific case involving edge states in a quantum Hall
liquid for discussing dissipation in a junction. In such systems, currents
flow only along the edges as all states in the bulk are localized; such edge 
modes are chiral in nature and can be described by theories of chiral bosons
\cite{wen1}. In the linear response regime, if a voltage $V$ is applied to an
Ohmic contact which is assumed to be perfectly coupled to the edge, then for
filling fraction $\nu$, the current $J$ injected into the edge from that
contact is given by $J = G V$, where $G = \nu e^2 /h$ is the conductance.

Now let us derive an expression for the power dissipated by such a system which
is governed by a current splitting matrix $\Mm$. The power $P_d$ dissipated
near the junction is given by the difference of the total incoming and 
outgoing powers \cite{wen2,halperin},
\bea P_d &=& \frac{1}{2} ~\sum_i ~(~ J_{Ii} ~V_{Ii} ~-~ J_{Oi} ~V_{Oi} ~)
\non \\
&=& \frac{1}{2G} ~J_I^T ~(I ~-~ \Mm^T \Mm) ~J_I, \label{power} \eea
where we have introduced a matrix notation in the second line of Eq.
(\ref{power}), with $J_I$ being a column made up of the incoming currents
$J_{Ii}$, and $I$ being the $N \times N$ identity matrix.
On physical grounds, the power dissipated near the junction cannot
be negative. This implies that the eigenvalues $\la_i$ of $\Mm^T \Mm$
must necessarily lie in the range $[0,1]$. If the incoming current $J_I$
is proportional to an eigenvector of $\Mm^T \Mm$ with eigenvalue $\la_i$,
the power dissipated will be proportional to $1- \la_i$. The set of values
of $1- \la_i$ therefore provides a measure of the amount of dissipation
associated with a system characterized by the current splitting matrix $\Mm$.

We emphasize here that Eq. (\ref{power}) describes the power dissipated
in the region close to the junction, and not in the leads which are assumed 
to be far away from the junction. For instance, for a two-wire junction with 
perfect transmission of the currents, i.e., for a matrix $\Mm$ given by 
$M_{11} = M_{22} = 0$ and $M_{12} = M_{21} = 1$, the expression in Eq. 
(\ref{power}) vanishes; however, we know that dissipation occurs in the 
leads because the outgoing electrons eventually equilibrate to the chemical 
potential there, leading to a contact resistance of $e^2 /h$ (for spinless 
electrons). Thus, $M^T M = I$ only means that there is no dissipation 
associated with the junction, although dissipation can still occur in the 
leads.

Since each row and column of $\Mm$ adds up to unity, both $\Mm$ and $\Mm^T$
must have one eigenvalue equal to $1$,
the corresponding eigenvector being given by a column all of whose entries
are equal to each other. This column is therefore an eigenvector of $\Mm^T
\Mm$ with eigenvalue equal to $1$ which corresponds to a situation where the
bias voltages $V_{Ii}$ (or incoming currents $J_{Ii}$) on all the wires are
equal, and no power is dissipated. Also note that the power dissipated
vanishes for {\it all} possible values of the incoming currents if $\Mm$ is
orthogonal. On the other hand, the dissipated power is maximized if all the
eigenvalues of $\Mm^T \Mm$ are equal to 0 except for one eigenvalue which
is necessarily equal to $1$. This occurs when all the entries of $\Mm$
are equal to $1/N$. Hence if we think of a situation where the dissipation
happens at the junction and not in the leads, the entire incoming power will
be converted to heat at the junction and the outgoing power will vanish.

In general, a current splitting matrix $\Mm$ corresponding to an
$N$-wire junction has $(N-1)^2$ independent parameters. This is because the
first $(N-1) \times (N-1)$ block of $\Mm$ can have arbitrary entries while the
entries of the last row and column of $\Mm$
are then fixed by the conditions that each row and column must add up to 1.
For $N=2$, we need only one parameter and the matrix is given by
\beq \Mm ~=~ \left( \begin{array}{cc} a & 1 - a \\
1 - a & a \end{array} \right), \eeq
where $a$ must lie in the range $[0,1]$ to ensure that the dissipated power is
always non-negative. No power will be dissipated if $a=0$ or 1 (i.e., $\Mm$ is
orthogonal), while the maximum power can be dissipated if $a=1/2$. It is
interesting to note that this one parameter family of $\Mm$-matrices can be
obtained from the following electronic $\Sm$-matrix describing
scattering of non-interacting electrons,
\beq \Sm ~=~ \left( \begin{array}{cc} \pm\sqrt{a} & \sqrt{1 - a} \\
\sqrt{1 - a} & \mp \sqrt{a} \end{array} \right). \eeq
The case of maximum dissipation, \i.e, $a=1/2$ for the non-interacting
electrons case also corresponds to extremal shot noise \cite{buttiker2} as is
expected.

For $N=3$, we require four parameters to specify $\Mm$ in general as we
can see below
\beq \Mm ~=~ \left( \begin{array}{ccc} a & b & 1 - a - b \\
c & d & 1 - c - d \\
1 - a - c & 1 - b - d & a + b - c - d-1 \end{array} \right). \label{four} \eeq
(The ranges of the parameters $a-d$ are fixed by the condition that the power
dissipated must be non-negative; hence we will not specify these ranges 
here). If we demand that no power be dissipated, i.e., that $\Mm$ be
orthogonal, then we only need to specify one parameter as will be discussed
below. Note that the three-wire case is quite different from
the two-wire case discussed earlier. In the three-wire case, it was possible
for $\Mm$ to have some negative elements without violating current
conservation and non-negativity of the dissipated power, in sharp contrast
to the two-wire case. To get a better feel for this, let us
consider a one-parameter family of $\Mm$-matrices which corresponds to a
highly symmetric junction given by
\beq \Mm ~=~ \left( \begin{array}{ccc} a & (1-a)/2 & (1-a)/2 \\
(1-a)/2 & a & (1-a)/2 \\
(1-a)/2 & (1-a)/2 & a \end{array}\right). \label{symmetric} \eeq
This matrix corresponds to a situation in which the reflected current in
each wire and the transmitted currents from one wire to the other two are
the same for all the wires. Using the condition of non-negativity of the net
dissipated power, we can show that the parameter $a$ must lie between $-1/3$
and $1$. For $a=-1/3$ and 1, $\Mm$ is orthogonal
and is therefore dissipationless. For all values of $a$ lying between
$-1/3$ and $0$ the diagonal elements of $\Mm$ are negative.
It is easy to see that $\Mm$-matrices with negative entries cannot
be obtained from any unitary $\Sm$-matrix, i.e., cannot be obtained from
any non-interacting electron theory. Hence such $\Mm$-matrices necessarily
correspond to situations in which the inter-electron interaction strength
is non-zero. We emphasize that such current splitting matrices only exist for
a junction of three or more wires and are absent for the two-wire case.

\begin{figure}[t]
\begin{center} \includegraphics[width=.7\linewidth]{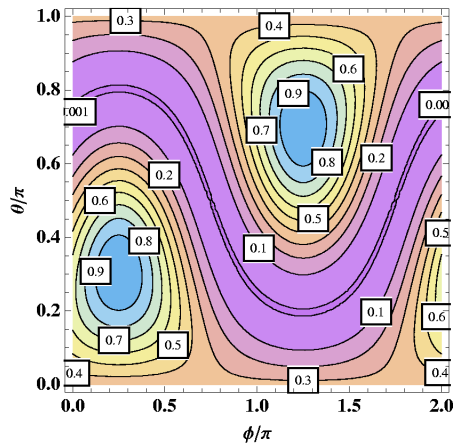} 
\end{center}
\caption{(Color online) Contour plot of $P_o$ given in Eq. (\ref{po}) as a 
function of $\phi$ and $\theta$.} \label{fig1} \end{figure}

Next, let us consider a situation where the power associated with the
incoming current is set to unity in units of $1/(2G)$. Then the
three-element column given by $J_I$ can be identified with a unit vector in
three dimensions which can be parametrized as $(\sin \theta \cos \phi,
\sin \theta \sin \phi, \cos \theta)$. The
maximum dissipation occurs for the $\Mm$-matrix which has all its
elements equal to $1/N$. This corresponds to $a=1/3$ in Eq.
(\ref{symmetric}). Using Eq. (\ref{power}) for the case of $a=1/3$, we
find that the power associated with the outgoing currents is given by
\beq P_o ~=~ \frac{1}{3} ~[cos \theta + sin \theta (cos \phi + sin \phi)]^2 .
\label{po} \eeq
Note that $P_o$ is bounded by $[0,1]$ and is symmetric under $\theta \to
\pi - \theta$ and $\phi \to \phi + \pi$. To visualize the expression in
Eq. (\ref{po}), we present a contour plot of $P_o$ as a function of $\phi$ 
and $\theta$ in Fig. \ref{fig1}. An interesting
point to note in the figure is the existence of a line of points on
which the outgoing power is zero and hence the power dissipation is maximum.
This implies that there is a family of bias voltage or incoming current
configurations for which the power dissipated is maximum. On the other hand,
there are two points at which the outgoing power is unity which correspond to
zero power dissipation. To understand these patterns, we recall that the
direction of maximum power dissipation in the space of incoming current
vectors corresponds to the two distinct eigenvectors of the $\Mm^T \Mm$ with
zero eigenvalue. For any value of $a$ in Eq. (\ref{symmetric}), an orthonormal
set of eigenvectors of $\Mm^T \Mm$ is given by ${\mathbf V_1}=(1,1,1)/\sqrt{3},
{\mathbf V_2}=(1,-1,1)/\sqrt{2},{\mathbf V_3}=(1,1,-2)/\sqrt{6}$ and the
corresponding eigenvalues are $1$, $(1-3 a)^2 /4$ and $(1-3 a)^2 /4$.
Note that these eigenvectors are independent of the parameter $a$. This
is so because of the symmetric form of $\Mm$ matrix. Hence, for this entire 
family of $\Mm$-matrices (Eq. (\ref{symmetric})), the eigenvectors (i.e.,
the combinations of incoming currents) which give the
maximum power dissipation are independent of $a$. Second,
the eigenvalue which is different from unity is a quadratic
function of $a$ which is zero for $a=1/3$ (maximum dissipation) and
unity for $a=-1/3$ and $a=1$ (both corresponding to zero dissipation). The
line of maximum dissipation appearing in Fig. \ref{fig1} corresponds to an
incoming current column $J_I$ which is a linear combination of the two
degenerate eigenvectors corresponding to the zero eigenvalue given by
$J_I=cos \de ~{\mathbf V_2} + sin \de ~{\mathbf V_3}$,
where $\de$ lies in the interval $[0,2\pi]$. The existence
of such a line of maximum dissipation is encouraging from an
experimental point of view since this implies that we only need to vary
a single parameter in an experiment to encounter the point of maximum
dissipation. The two points in Fig. \ref{fig1} which have $P_o =1$
(zero dissipation) correspond to the eigenvector ${\mathbf V_1}$. There are
two such points because we get zero dissipation if the incoming current is
prepared either in the direction of this eigenvector or opposite to it;
in Fig. \ref{fig1}, these points lie at $(\phi,\theta) = (\pi/4, \cos^{-1} (1/
\sqrt{3}))$ and $(5\pi/4, \pi - \cos^{-1} (1/ \sqrt{3}))$.
To conclude, we see that a study of the eigenvectors of $\Mm^T \Mm$ can lead
to a complete understanding of dissipation in a junction as a function of
the bias voltages applied in the various wires.

\section{A three-wire model with dissipation}

\begin{figure}[htb]
\begin{center} \includegraphics[width=.7\linewidth]{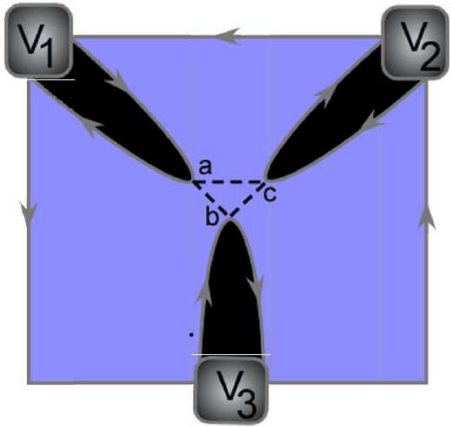} 
\end{center}
\caption{(Color online) Picture of a three-wire model with tunneling 
conductances $\si_{ij}$ between points $i,j$ which can take values $a,b,c$. 
$V_i$ denote the incoming voltages.} \label{fig2} \end{figure}

In this section, we develop a microscopic model for a three-wire system
with a dissipative junction. A schematic picture of the system is presented in
Fig. \ref{fig2}. The currents and voltages on each wire will be assumed to be
governed by $J = G V$ (where $G = \nu e^2 /h$) on all the incoming and 
outgoing chiral wires. (The symbols $V_i$ in the figure denote the incoming 
voltages which drive the incoming currents; the outgoing currents and voltages
are then determined by the $V_i$ and the matrix $\Mm$ which will be derived
below). The junction region consists of three
points $a,b,c$, one point lying on each of the three wires as shown in
Fig. \ref{fig2}. Electrons can tunnel between any two of these points, say,
$i$ and $j$. If the tunneling amplitude is denoted by $\xi_{ij}$, the
corresponding tunneling conductance $\si_{ij} G$ will be proportional to
$|\xi_{ij}|^2$. Here we have introduced the quantity $G$ so that
$\si_{ij}$ is dimensionless. The conductances satisfy $\si_{ij} = \si_{ji}
\ge 0$. If the voltages at the two points are given by $V_i$ and $V_j$,
the current flowing from $i$ to $j$ will be given by $\si_{ij} G (V_i - V_j)$.
We will now see that this model gives rise to a current splitting matrix $\Mm$
which is generally dissipative.

In our analysis, we will work directly with the currents without introducing
any fermionic or bosonic fields. To derive the matrix $\Mm$, we have to
determine the outgoing currents $(J_{O1}, J_{O2}, J_{O3})$ in terms of the
incoming currents $(J_{I1}, J_{I2}, J_{I3})$. The incoming and outgoing
currents (and therefore voltages) will generally change discontinuously at
the three junction points. The corresponding incoming and outgoing voltages
are obtained by dividing the currents by $G$. We assume that the voltages at
each of the three points of the junction are given by the mean values of the
corresponding incoming and outgoing voltages. Namely,
\beq V_i ~=~ \frac{1}{2}~(~V_{Ii} ~+~ V_{Oi}~) ~=~ \frac{1}{2G} ~(~J_{Ii} ~
+~ J_{Oi}~) \label{mean} \eeq
for $i=1,2,3$.

The mean value assumption made in Eq. (\ref{mean}) can be justified as
follows. We can begin with a model in which the tunneling region is not a
point but has a finite length $l$, and there is a tunneling conductance per
unit length given by $\tilde \si_{ij}$ \cite{agarwal1,kane1}. Tunneling will
then occur from every point lying in the tunneling
region in wire $i$ to the corresponding point lying in wire $j$. We then
find that the current $J_i$ on wire $i$ changes smoothly from $J_{Ii}$ to
$J_{Oi}$ as we go from one end of the tunneling region to the other. Hence
the voltage $V_i (x_i) = J_i (x_i) /G$ will also change smoothly, where $i$
runs over $1,2,3$, and $x_i$ runs over the tunneling region from 0 to $l$.
The current $J_i (x_i)$ can be obtained by solving equations of continuity
given by \cite{agarwal1,kane1}
\beq \frac{\pa J_i}{\pa x_i} ~=~ - ~\sum_{j \ne i} ~{\tilde \si}_{ij} ~[J_i
(x_i) - J_j (x_j)], \eeq
We can solve these equations to obtain the dependence of the
current $J_i (x_i)$ and voltage $V_i (x_i) = J_i (x_i)/G$ on the coordinates
$x_i$. If we now take the limit $l \to 0$ with $l {\tilde \si}_{ij} = \si_{ij}$
being held fixed, we recover the earlier model of tunneling between three
points, with the voltages at the three points being given by Eq. (\ref{mean}).

We now return to our original model and write down equations of continuity for
the currents at the three tunneling points,
\bea J_{Oi} ~-~ J_{Ii} ~=~ -~ G ~\sum_{j \ne i} ~\si_{ij} ~(V_i ~-~ V_j),
\label{vij} \eea
for $i=1,2,3$. Using Eq. (\ref{mean}), we can solve for the $J_{Oi}$ in
terms of the $J_{Ii}$. This enables us to obtain the matrix $\Mm$ which relates
the two sets of currents. We find that
\bea \Mm_{11} &=& \frac{1 ~+~ \si_{23} - (1/4) S_2}{1 ~+~ S_1 ~+~ (3/4) S_2},
\non \\
\Mm_{12} &=& \frac{\si_{12} ~+~ (1/2) S_2}{1 ~+~ S_1 ~+~ (3/4) S_2},
\label{mat} \eea
where $S_1 \equiv \si_{12} + \si_{23} +\si_{31}$ and $S_2 \equiv \si_{12}
\si_{23} + \si_{23} \si_{31} + \si_{31} \si_{12}$. All the other entries
of $\Mm$ can be found by symmetry. Note that each row and column of $\Mm$
adds up to 1 as desired. In addition, $\Mm$ being a symmetric matrix 
is a special feature of this specific model.

We can show in general that the expression for power dissipation given in
Eq. (\ref{power}) agrees with the sum of the powers dissipated by the three
tunneling processes. Namely, if we substitute the expression for $\Mm$ given 
in Eq. (\ref{mat}) in Eq. (\ref{power}), and compare that with the expression
for the power dissipated ($=I \times V$) by the three tunnelings, namely,
\beq G [\si_{12} ~(V_1 - V_2)^2 + \si_{23} ~(V_2 - V_3)^2 + \si_{31} (V_3 -
V_1)^2], \eeq
(where $V_i - V_j$ appears in Eq. (\ref{vij})), we find that the two agree
for all values of the incoming currents $J_{Ii}$.

In the special case that $\si_{12} = \si_{23} = \si_{31} = \si$, the
expression for $\Mm$ simplifies to
\beq \Mm ~=~ \frac{1}{1 + 3\si /2} ~\left( \begin{array}{ccc}
1 - \si /2 & \si & \si \\
\si & 1 - \si /2 & \si \\
\si & \si & 1 - \si /2 \end{array} \right), \label{siM} \eeq
which is of the form given in Eq. (\ref{symmetric}).
Three particular values of this $\Mm$ are worth noting, namely,
\bea \Mm &=& \left( \begin{array}{ccc}
1 & 0 & 0 \\
0 & 1 & 0 \\
0 & 0 & 1 \end{array} \right) ~~{\rm for}~~ \si ~=~ 0, \non \\
&=& \left( \begin{array}{ccc}
1/3 & 1/3 & 1/3 \\
1/3 & 1/3 & 1/3 \\
1/3 & 1/3 & 1/3 \end{array} \right) ~~{\rm for}~~ \si ~=~ 2/3, \non \\
&=& \left( \begin{array}{ccc}
-1/3 & 2/3 & 2/3 \\
2/3 & -1/3 & 2/3 \\
2/3 & 2/3 & -1/3 \end{array} \right) ~~{\rm for}~~ \si ~=~ \infty.
\label{mmatrix} \eea

For the $\Mm$-matrix given in Eq. (\ref{siM}), one of the eigenvalues of
$\Mm^T \Mm$ is equal to 1, while the other two are equal to
$[(2-3\si)/(2+3\si)]^2$. There
is no power dissipation ($\Mm$ is orthogonal) if $\si = 0$
or $\infty$, while there is maximum power dissipation if $\si = 2/3$. On
physical grounds it is natural to expect that there is no dissipation
if $\si = 0$. But the dissipation also turns out to to be zero for
$\si = \infty$ which is somewhat surprising. This can be traced back to the
analysis done in Ref. \onlinecite{chamon} for the junction of three TLL
wires. In that paper, the authors started with a situation where
there is a perfectly reflecting (disconnected) junction of three TLL wires
effectively described by a $\Mm$-matrix corresponding to $\si = 0$
in Eq. (\ref{siM}), and then
switched on electron tunneling operators between each pair of wires
such that the amplitudes of all the three tunneling operators are equal.
Using the technique of bosonization, they then established that as the
strengths of all the tunneling operators go to infinity under an RG
flow, the system is described by the $\Mm$-matrix which is obtained
by taking the $\si = \infty$ limit in Eq. (\ref{siM}). In their analysis,
the $\Mm$-matrices
corresponding to both $\si = 0$ and $\si = \infty$ are fixed points of the
theory, and they are connected to each other by a duality transformation.
These statements make our model seem quite attractive because even
though it is rather simple, it manages to capture the essential non-trivial
physics related to dissipation in a three-wire junction without getting into
the technicalities of bosonization.

If the various edges shown in Fig. \ref{fig2} are the edges of a quantum
Hall system, the tunneling operators will satisfy some RG equations. Depending
on the filling fraction $\nu$ and the location of the quantum Hall liquid
with respect to the edges \cite{das2006}, the tunneling operators will be
either irrelevant or relevant, and the corresponding tunneling conductances
will then flow to 0 or $\infty$ respectively. This implies that the RG fixed 
point for $\Mm$ will be given by either the first matrix or the last matrix 
in Eq. (\ref{mmatrix}). Thus, the second matrix in (\ref{mmatrix}) which 
corresponds to maximum dissipation does not appear to be a fixed point of 
the model.

\section{A more complex three-wire model}

We now consider another model for a dissipative junction of three wires.
This model consists of a ring shaped region (with two chiral edges) and three
external wires (each with two chiral edges: incoming and outgoing) which
connect to the ring at three different points. All the edges carry currents
and can be modeled by TLLs. Further, each of the external wires can have
different bias voltages which determine the incoming currents impinging on the
ring region. Along the ring, the co-propagating currents can tunnel between 
the two edges. For simplicity we will assume equal tunneling amplitudes at 
all points.

\begin{figure}[htb]
\includegraphics[width=.8\linewidth]{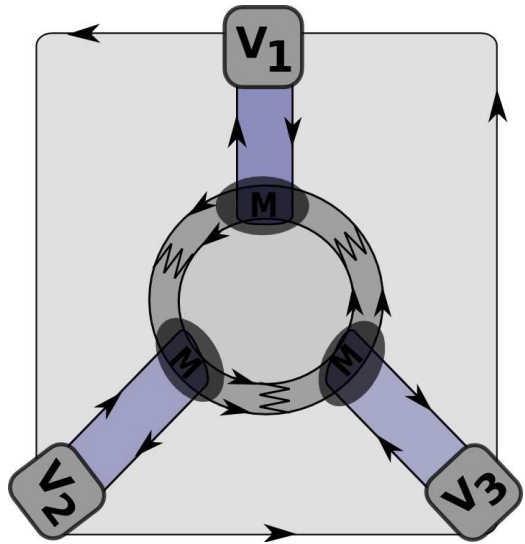}
\caption{(Color online) Picture of a three-wire model where the external 
wires connect to a ring which has two co-propagating edges with interedge 
tunneling. $V_i$ denote the incoming voltages.} \label{fig3} \end{figure}

Fig. \ref{fig3} gives a schematic picture of the model we have in mind.
Each of the external wires is made of a TLL (or a single edge of a fractional
quantum Hall system) and has an outgoing and an incoming chiral edge $I_{Oi}$
and $I_{Ii}$, where $i$ labels the wire. The ring is also made of a TLL
(or a single mode fractional quantum Hall edge) and has two co-propagating
modes, one on the outer edge ($J_{Oi}$) and the other on the inner edge
($J_{Ii}$).

At each of the `point' junctions where an external wire meets the ring, we
have three incoming modes and three outgoing modes (marked by arrows in Fig.
\ref{fig3}). Such a three-wire junction can be described by a orthogonal
$3 \times 3$ current splitting matrix ($\Mm$) whose rows and columns add
up to 1. The orthogonality implies that (i) the junction relates the outgoing
bosonic fields to the incoming bosonic fields in a way which preserves
the chiral commutation relations of the fields, and (ii) the
junction is dissipationless. For a three-wire charge-conserving and
dissipationless junction, the matrix $\Mm$ can be parametrized by a single
continuous parameter $\theta$ \cite{chamon,das2006,agarwal2},
and it can be classified into two classes for which (a) $\det \Mm_1 = 1$,
and (b) $\det \Mm_2=-1$. These two classes are expressed as
\beq \label{m1} \Mm_1 = \left(\begin{array}{ccc}
a & b & c \\
c & a & b \\
b & c & a \end{array}\right), \quad \Mm_2 = \left(\begin{array}{ccc}
b & a & c \\
a & c & b \\
c & b & a \end{array}\right). \eeq
In Eq. (\ref{m1}), $a=(1+2\cos\theta)/3$, $b=(1-\cos \theta + \sqrt{3} \sin
\theta)/3$, and $c=(1-\cos\theta -\sqrt{3} \sin\theta)/ 3$.
In the $\Mm_1$ class, $\theta = 0$ corresponds to the disconnected $N$ fixed
point, $\theta= \pi$ to the $D_P$ fixed point, and $\theta = \pm 2\pi/3$ to
the chiral fixed points $\chi_{\pm}$ in the notation of Ref. 
\onlinecite{chamon}.

Each of the three `point' junctions in Fig. \ref{fig3} is characterized by a
dissipationless current splitting matrix $\Mm$. For simplicity we will now
assume all the three junctions have the same $\Mm$ with the same orientation.
We will also assume all the `point' junction matrices to be identical and of
the $\Mm_1$ type. Next, we will allow tunneling between
the inner and outer edges of the ring, which can be thought of `classically'
as a resistor connecting the inner and outer current carrying wires. A more 
microscopic model of such a dissipative tunneling is given in Refs. 
\onlinecite{agarwal1} and \onlinecite{kane1}. The main result is that at the 
ends of each tunneling region (of length $L$), the currents on the outgoing 
edges are a linear combination of the incoming currents and can be written as
\beq \left(\begin{array}{c} J_{O1}(L) \\ J_{O2}(L) \end{array}\right) ~=~
\left( \begin{array}{cc} 1 - t &
t \\ t & 1- t\end{array} \right) \left(\begin{array}{c} J_{I1}(0) \\
J_{I2}(0) \end{array}\right). \eeq
The parameter $t$ can be expressed in terms of the microscopic tunneling
conductance as
\beq t ~=~ \frac{1}{2} ~(1-e^{-2L\si h/(\nu e^2)}) \label{t} \eeq
for the case of co-propagating edges, where $\si$ is the tunneling
conductance per unit length between the two edges of the ring \cite{agarwal1}.

We note again that the `point' junction matrices connecting external wires to
the ring are dissipationless. The only source of dissipation in our model is
therefore the interedge tunneling between the co-propagating modes 
propagating on the ring.
Now, starting from a given dissipationless current splitting matrix $\Mm_1$ at
each `point' junction and a given interedge tunneling parameter $t$, we can
solve for the three outgoing and six interedge currents in terms of the three
incoming currents. We then find that the $\Mm$-matrix of the system which
relates the outgoing currents to the incoming currents is of the cyclic form

\beq \left(\begin{array}{c} I_{O1} \\ I_{O2} \\ I_{O3} \end{array}\right)
= \left(\begin{array}{ccc}
d & e & f \\
f & d & e \\
e & f & d \end{array}\right) \left(\begin{array}{c}
I_{I1} \\ I_{I2} \\ I_{I3} \end{array}\right), \label{Md1} \eeq
where $d$, $e$ and $f$ are given by
\begin{widetext}
\bea d &=& \frac{30 t^2-48 t+ 27 + \left(60 t^2-84 t+42\right) \cos
\theta + \left(18 t^2-30 t+12\right) \cos (2 \theta)}{42 t^2-48 t + 33
+\left(28 t^2-68 t+34\right) \cos \theta +\left(38 t^2-46 t+14\right)
\cos (2 \theta)}, \non \\
e &=& \frac{12 t^2 + 6 t - \left(24 t^2-12 t+6\right) \cos \theta + \left(
12t^2 -18 t + 6 \right) \cos (2 \theta)}{42 t^2-48 t+ 33+ \left(28 t^2-68 t
+34 \right) \cos \theta +\left(38 t^2-46 t+14\right) \cos (2 \theta)}, \non \\
f &=& \frac{-6 t +6 - \left(8 t^2 -4t+ 2\right) \cos \theta + \left( 8t^2
+2t -4 \right) \cos (2 \theta)}{42 t^2-48 t+ 33 + \left(28 t^2-68 t+34 \right)
\cos \theta +\left(38 t^2-46 t+14\right) \cos (2 \theta)}. \label{def} \eea
If we take all the matrices at the three `point' junctions to be identical
and of the $\Mm_2$ type, we again find that the $\Mm$-matrix of the complete
system is of the cyclic form given in Eq. (\ref{Md1}), although the
expressions for $d,e,f$ are different from those given in Eq. (\ref{def}).
For the $\Mm$-matrix given in Eqs. (\ref{Md1}-\ref{def}), one of the
eigenvalues of $\Mm^T \Mm$ is equal to $1$ (non-dissipative),
while the other two (degenerate and dissipative) are given by
\beq \la ~=~ \frac{78 t^2-84 t+ 33 + \left(28 t^2-68 t+34\right) \cos \theta
+\left( 2t^2-10 t+14\right) \cos (2 \theta)}{42 t^2-48 t+ 33 + \left(28 t^2-68t
+ 34\right) \cos \theta + \left(38 t^2-46 t+14\right) \cos (2 \theta)}.
\label{la} \eeq
\end{widetext}

\begin{figure}[htb]
\includegraphics[width=.7\linewidth]{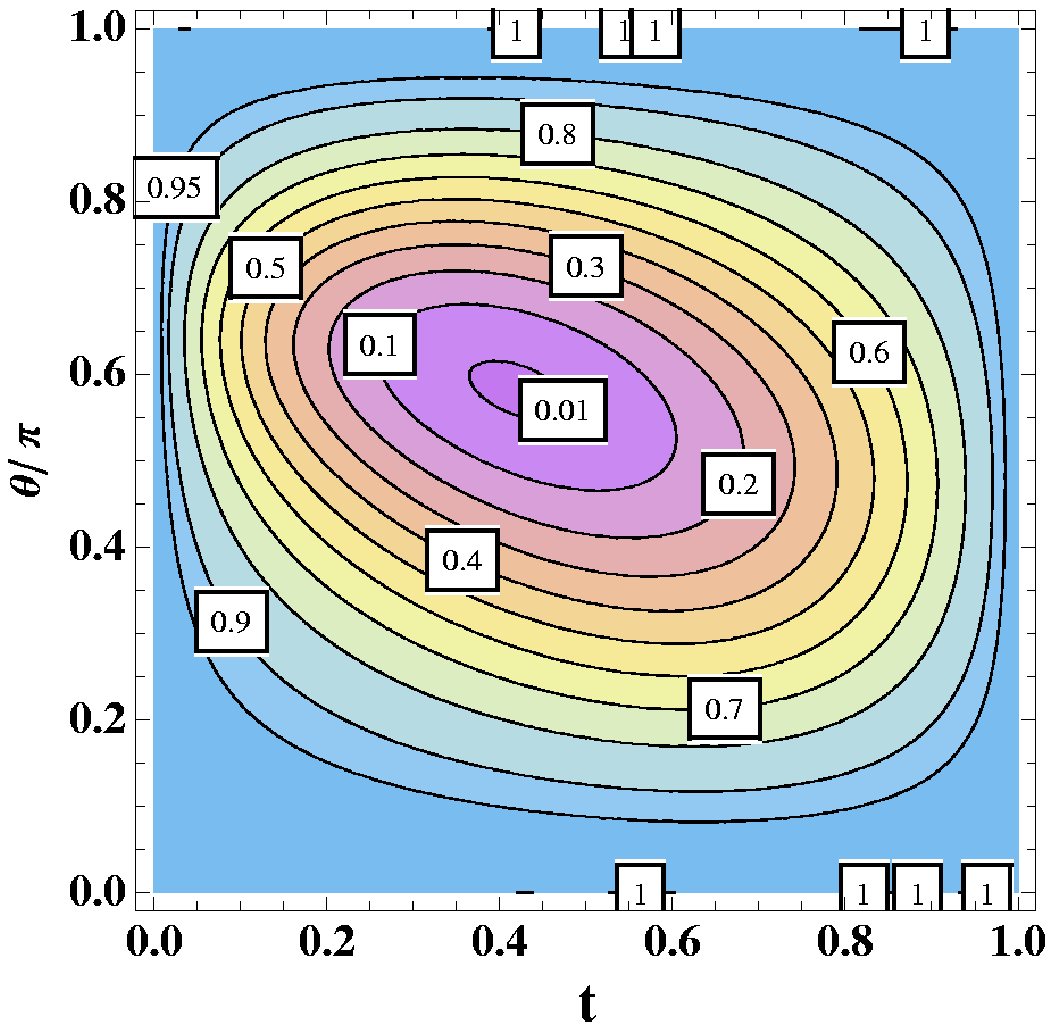}
\caption{(Color online) Contour plot of $\la$ given in Eq. (\ref{la}) as a 
function of $t$ and $\theta$.} \label{fig4} \end{figure}

A contour plot of $\la$ in the $t - \theta $ plane is presented in Fig.
\ref{fig4}. Since $\la$ is symmetric under $\theta \to - \theta$, we have
only plotted $\theta$ from 0 to $\pi$ in the figure. We see that $\la = 1$
(no dissipation) if either $\theta = 0, \pi$ or $t=0$. It was shown in Ref.
\onlinecite{agarwal1} that an RG flow takes the variable $L\si$ to either 0
or $\infty$, depending on the value of the interaction parameter of the TLLs
which constitute the two edges of the ring. Hence the fixed-point values of 
the parameter $t$ are 0 and 1/2 according to Eq. (\ref{t}). For $t \to 0$, the
eigenvalue $\la$ goes to 1 for any value of $\theta$ and we therefore get a
dissipationless $\Mm$-matrix. But for $t \to 1/2$, we find that
\beq \la ~=~ \frac{21+ 14 \cos \theta +19 \cos (2 \theta)}{39+ 14 \cos
\theta +\cos (2 \theta)}. \eeq
This is equal to 1 for $\theta = 0,\pi$, and is not equal to 0 for
any value of $\theta$. According to Fig. \ref{fig4},
the point of maximum dissipation ($\la =0$) lies at $(t,\theta /\pi)
\simeq (0.419,0.583)$, and not at $t=0$ or 1/2, and it is therefore not a
fixed point of this model.

\section{Discussion and Summary}

In this work we introduced a scheme to quantify dissipation for a $N$-wire
junction for both non-interacting electrons and TLL wires. The quantification
is achieved in terms of a real current splitting matrix $\Mm$. The
dissipated power can be parametrized by the non-zero eigenvalues of $I -
\Mm^T \Mm$ and hence there is no dissipation if $\Mm$ is orthogonal since
$I - \Mm^T \Mm$ is then equal to 0. We have shown that if an
eigenvalue of $I - \Mm^T \Mm$ is equal to 1, the corresponding eigenvector
determines a combination of the applied bias voltages for which the input
power is completely dissipated at the junction. For a three-wire junction,
the matrix $\Mm$ with all entries equal to $1/3$ has a doubly degenerate
eigenvalue equal to 1. Hence any linear combination of the two eigenvectors 
corresponds to a combination of bias voltages which will lead to complete 
dissipation at the junction. This implies that the bias voltage
combination which corresponds to maximum dissipation is not a unique point 
in the allowed parameter space but forms a one-parameter family of points as 
discussed in Sec. III. This fact makes it more likely to be accessible in an 
experimental situation.

We presented two microscopic models of dissipation for a three-wire
system, one involving tunneling between three points (Sec. IV) and the other
involving tunneling between three pairs of edges lying on a ring (Sec. V).
The model in Sec. IV leads to a symmetric $\Mm$-matrix depending
on three parameters $\si_{ij}$, while the model in Sec. V leads, for a
particular choice of current splitting matrices at the `point' junctions, to
a cyclic $\Mm$-matrix depending on two parameters $t,\theta$.

For both models, we have briefly discussed the RG flows of the various
parameters. For the model in Sec. IV, the RG flow takes the system to one of
two fixed points, both of which correspond to dissipationless $\Mm$-matrices.
For the model in Sec. V, the RG flow again takes the system to one of two
fixed points, one of which gives a dissipationless matrix while the other
is generally dissipative (except for the special cases $\theta = 0, \pi$).
In all cases, we find that the matrix corresponding to maximum dissipation
(i.e., all elements of $\Mm$ being equal to 1/3) is not a fixed point of the
RG equations. Hence within the models we have studied, it appears that there
is nothing special about the maximally dissipative $\Mm$-matrix from an
RG point of view.

\section*{Acknowledgments}

We thank Sumathi Rao for interesting discussions, and DST, India for
financial support under Project No. SR/S2/CMP-27/2006.


\begin{thebibliography}{99}

\bib{gogolin} A. O. Gogolin, A. A. Nersesyan, and A. M. Tsvelik, {\it
Bosonization and Strongly Correlated Systems} (Cambridge University Press,
Cambridge, 1998).

\bib{vondelft} J. von Delft and H. Schoeller, Ann. Phys. (Leipzig) {\bf 7},
225 (1998).

\bib{rao} S. Rao and D. Sen, in {\it Field Theories in Condensed Matter
Physics}, edited by S. Rao (Hindustan Book Agency, New Delhi, 2001).

\bib{giamarchi1} T. Giamarchi, {\it Quantum Physics in One Dimension} (Oxford
University Press, Oxford, 2004).

\bib{bruus} H. Bruus and K. Flensberg, {\it Many-Body Quantum Theory in
Condensed Matter Physics: An Introduction} (Oxford University Press, Oxford,
2004).

\bib{giuliani} G. F. Giuliani and G. Vignale, {\it Quantum Theory of Electron
Liquid} (Cambridge University Press, 2005).

\bib{li} J. Li, C. Papadopoulos, and J. Xu, Nature (London) {\bf 402}, 253 
(1999).

\bib{kumar} B. C. Satishkumar, P. J. Thomas, A. Govindaraj, and C. N. R. Rao,
Appl. Phys. Lett. {\bf 77}, 2530 (2000).

\bibitem{papad} C. Papadopoulos, A. Rakitin, J. Li, A. S. Vedeneev, and J. M.
Xu, \prl{\bf 85}, 3476 (2000).

\bib{fuhrer} M. S. Fuhrer, J. Nygard, L. Shih, M. Forero, Y.-G. Yoon, M. S.
C. Mazzoni, H. J. Choi, J. Ihm, S. G. Louie, A. Zettl, and P. L. McEuen,
Science {\bf 288}, 494 (2000).

\bib{terrones} M. Terrones, F. Banhart, N. Grobert, J.-C. Charlier, H.
Terrones, and P. M. Ajayan, Phys. Rev. Lett. {\bf 89}, 075505 (2002).

\bib{sandler} N. P. Sandler, C. C. Chamon, and E. Fradkin, Phys. Rev. B
{\bf 57}, 12324 (1998), and Phys. Rev. B {\bf 59}, 12521 (1999).

\bib{nayak} C. Nayak, M. P. A. Fisher, A. W. W. Ludwig, and H. H. Lin,
\prb{\bf 59}, 15694 (1999).

\bib{lal1} S. Lal, S. Rao, and D. Sen, \prb{\bf 66}, 165327 (2002).

\bib{chen} S. Chen, B. Trauzettel, and R. Egger, \prl{\bf 89}, 226404 (2002);
R. Egger, B. Trauzettel, S. Chen, and F. Siano, New J. Phys. {\bf 5}, 117 
(2003).

\bib{chamon} C. Chamon, M. Oshikawa, and I. Affleck, \prl{\bf 91}, 206403
(2003); M. Oshikawa, C. Chamon, and I. Affleck, J. Stat. Mech.: Theory Exp.
{\bf 0602}, P008 (2006).

\bib{meden} X. Barnabe-Theriault, A. Sedeki, V. Meden, and K. Sch\"onhammer,
\prb{\bf 71}, 205327 (2005), and \prl{\bf 94}, 136405 (2005).

\bib{das2006} S. Das, S. Rao, and D. Sen, Phys. Rev. B {\bf 74}, 045322 (2006).

\bib{lal2} S. Lal, Phys. Rev. B {\bf 77}, 035331 (2008).

\bib{rosenow} B. Rosenow and B. I. Halperin, arXiv:0806.0869v2 (unpublished).

\bib{giuliano} D. Giuliano and P. Sodano, Nucl. Phys. B {\bf 811}, 395 (2009),
and New J. Phys. {\bf 10}, 093023 (2008).

\bib{bellazzini} B. Bellazzini, M. Burrello, M. Mintchev, and P. Sorba,
arXiv:0801.2852; B. Bellazzini, P. Calabrese, and M. Mintchev, Phys. Rev. B
{\bf 79}, 085122 (2009).

\bib{das2007} S. Das, S. Rao, and A. Saha, EPL {\bf 81}, 67001 (2008), and 
Phys. Rev. B {\bf 77}, 155418 (2008).

\bib{das2008} S. Das and S. Rao, Phys. Rev. B {\bf 78}, 205421 (2008).

\bib{linejunction} C. L. Kane and M. P. A. Fisher, Phys. Rev. B {\bf 56}, 
15231 (1997).

\bib{buttiker1} M. B\"uttiker, Phys. Rev. B {\bf 46}, 12485 (1992).

\bib{agarwal1} D. Sen and A. Agarwal, Phys. Rev. B {\bf 78}, 085430 (2008).

\bib{agarwal2} A. Agarwal, S. Das, S. Rao, and D. Sen, Phys. Rev. Lett.
{\bf 103}, 026401 (2009).

\bib{wen1} X.-G. Wen, Phys. Rev. Lett. {\bf 64}, 2206 (1990), and Int. J. Mod.
Phys. B {\bf 6}, 1711 (1992).

\bib{wen2} X.-G. Wen, Phys. Rev. B {\bf 50}, 5420 (1994).

\bib{halperin} D. B. Chklovskii and B. I. Halperin, Phys. Rev. B {\bf 57},
3781 (1998).

\bib{buttiker2} Ya M. Blanter and M. B\"uttiker, Phys. Rep. {\bf 336}, 1 
(2000).

\bib{kane1} C. L. Kane and M. P. A. Fisher, Phys. Rev. B {\bf 52}, 17393
(1995). 

\end{thebibliography}
\end{document}